\documentclass[twocolumn]{aastex631}
\hypersetup{linkcolor=blue, citecolor=cyan, filecolor=cyan, urlcolor=blue}
\usepackage{multirow}
\usepackage{amsmath}
\usepackage[figuresright]{rotating}

\shorttitle{ Non-parametric Morphology of Galaxies in {\it JWST}}
\shortauthors{J. Ren et al.}
\graphicspath{{./}{figures/}}

\begin{document}

%\correspondingauthor{XXX}
\email{E-mail: fsliu@nao.cas.cn, nan.li@nao.cas.cn}

 \title{Calibrating non-parametric morphological indicators from  {\it JWST} images for galaxies over $0.5<z<3$}

\author [0000-0002-5043-2886]{Jian Ren}
\affil{National Astronomical Observatories, Chinese Academy of Sciences, 20A Datun Road, Chaoyang District, Beijing 100101, China}
\affil{Key Laboratory of Space Astronomy and Technology, National Astronomical Observatories, Chinese Academy of Sciences, 20A Datun Road, Chaoyang District, Beijing 100101, China}

\author{F. S. Liu $^{\color{blue} \dagger}$}
\affil{National Astronomical Observatories, Chinese Academy of Sciences, 20A Datun Road, Chaoyang District, Beijing 100101, China}
\affil{Key Laboratory of Optical Astronomy, National Astronomical Observatories, Chinese Academy of Sciences, 20A Datun Road, Chaoyang District, Beijing 100101, China}
\affil{School of Astronomy and Space Science, University of Chinese Academy of Sciences, Beĳing 100049, China}

\author{Nan Li $^{\color{blue} \dagger}$}
\affil{National Astronomical Observatories, Chinese Academy of Sciences, 20A Datun Road, Chaoyang District, Beijing 100101, China}

\affil{Key Laboratory of Space Astronomy and Technology, National Astronomical Observatories, Chinese Academy of Sciences, 20A Datun Road, Chaoyang District, Beijing 100101, China}
\affil{School of Astronomy and Space Science, University of Chinese Academy of Sciences, Beĳing 100049, China}

\author{Qifan Cui}
\affil{National Astronomical Observatories, Chinese Academy of Sciences, 20A Datun Road, Chaoyang District, Beijing 100101, China}
\affil{Key Laboratory of Space Astronomy and Technology, National Astronomical Observatories, Chinese Academy of Sciences, 20A Datun Road, Chaoyang District, Beijing 100101, China}

\author{Pinsong Zhao}
\affil{National Astronomical Observatories, Chinese Academy of Sciences, 20A Datun Road, Chaoyang District, Beijing 100101, China}
\affil{Key Laboratory of Optical Astronomy, National Astronomical Observatories, Chinese Academy of Sciences, 20A Datun Road, Chaoyang District, Beijing 100101, China}

\author{Yubin Li}
\affil{National Astronomical Observatories, Chinese Academy of Sciences, 20A Datun Road, Chaoyang District, Beijing 100101, China}
\affil{Key Laboratory of Space Astronomy and Technology, National Astronomical Observatories, Chinese Academy of Sciences, 20A Datun Road, Chaoyang District, Beijing 100101, China}

\author{Qi Song}
\affil{National Astronomical Observatories, Chinese Academy of Sciences, 20A Datun Road, Chaoyang District, Beijing 100101, China}

\author{Hassen M. Yesuf}
\affil{
Key Laboratory for Research in Galaxies and Cosmology, Shanghai Astronomical Observatory, Chinese Academy of Sciences, 80 Nandan Road, Shanghai 200030, China}

\author{Xian Zhong Zheng}
\affil{Purple Mountain Observatory, Chinese Academy of Sciences, 10 Yuanhua Road, Nanjing 210023, China}
\affil{School of Astronomy and Space Science, University of Science and Technology of China, Hefei 230026, China}

\begin{abstract}

The measurements of morphological indicators of galaxies are often influenced by a series of observational effects.  In this study, we utilize a sample of over 800 TNG50 simulated galaxies with log($M_*$/M$_\odot$)$>9$ at $0.5<z<3$ to investigate the differences in non-parametric morphological indicators ($C$, $S$, $Gini$, $M_{\rm 20}$, $A_{\rm O}$, and $D_{\rm O}$) derived from noise-free and high-resolution TNG50 images and mock images simulated to have the same observational conditions as {\it JWST}/NIRCam. We quantify the relationship between intrinsic and observed values of the morphological indicators and accordingly apply this calibration to over 4600 galaxies in the same stellar mass and redshift ranges observed in {\it JWST} CEERS and JADES surveys.  We find a significant evolution of morphological indicators with rest-frame wavelength ($\lambda_{\rm rf}$) at $\lambda_{\rm rf}<1$\,$\mu$m, while essentially no obvious variations occur at $\lambda_{\rm rf}>1$\,$\mu$m.  The morphological indicators of star-forming galaxies (SFGs) and quiescent galaxies (QGs) are significantly different. The morphologies of QGs exhibit a higher sensitivity to rest-frame wavelength than SFGs. After analyzing the evolution of morphological indicators in the rest-frame V-band (0.5-0.7\,$\mu$m) and rest-frame J-band (1.1-1.4\,$\mu$m), we find that the morphologies of QGs evolve substantially with both redshift and stellar mass. For SFGs, the $C$, $Gini$ and $M_{\rm 20}$ show a rapid evolution with stellar mass  at log($M_*$/M$_\odot$)$\geq10.5$, while the $A_{\rm O}$, $D_{\rm O}$ and $A$ evolve with both redshift and stellar mass. Our comparison shows that TNG50 simulations effectively reproduce the morphological indicators we measured from {\it JWST} observations when the impact of dust attenuation is considered.

\end{abstract}


\section{Introduction} \label{sec:intro}

The origin of the Hubble sequence of galaxy morphologies \citep{Hubble1926}, in connection with other fundamental aspects of galaxy evolution, has become a focal point in the study of galaxy formation and evolution. These aspects include stellar kinematics \citep{Rodrigues2017,Thob2019,Wang2020}, star formation processes \citep{Kennicutt1998,Martig2009,Lee2013,Yesuf2021}, AGN activity \citep{Kocevski2012,Gabor2009,Povic2012,Yesuf2020}, as well as the distribution of dust and stellar populations \citep{Wuyts2012}. Moreover, morphology is linked to the environments of galaxies \citep{Dressler1980, Dressler1997} and the prevalence of mergers and interactions \citep[e.g.][]{Toomre1972, Barnes1992, Bournaud2005}.

Based on Hubble Space Telescope ({\it HST}) observations, galaxy morphology at $z \sim 2$ was already discernible according to the Hubble sequence \citep{VandenBergh1996}. Moreover, the prevalence of different Hubble types was found to evolve drastically with redshift \citep{Ravindranath2004, Buitrago2008, Buitrago2013, Whitney2021}. Beyond $z=2$, galaxies exhibited primarily irregular morphologies \citep{Conselice2005, Mortlock2013}, with the more clumpy and irregular appearance in the rest-frame UV band resulting mainly in the {\it HST}/WFC3 F160W band \citep{Guo2015}. However, the recent multi-band images from the {\it JWST}/NIRCam have dramatically altered our understanding of the galaxy morphology of distant galaxies. An increasing body of evidence now suggests that these galaxies are predominantly disk-dominated \citep{Ferreira2022, Ferreira2023, Kartaltepe2023}, and remarkably, the Hubble sequence is observable even at $z=6$ \citep{Huertas-Company2023}. These most recent findings underscore the necessity to reassess the evolutionary trend in morphological structures of galaxies across different rest-frame wavelengths and redshifts.

Non-parametric analysis is widely utilized in the study of galactic morphology as it provides a direct measure of form without making assumptions about the galaxy structure \citep{Conselice2003a}. It has been effectively utilized in various research contexts, including morphological classification, the study of morphological evolution, and the investigation of the relationship between galaxy morphology and its intrinsic physical properties \citep{Conselice2014}. Non-parametric morphological indicators have revolutionized the analysis of galaxy structures by providing a simple, objective, and quantitative means of describing them, representing a significant advantage over other methods such as the visual Hubble classification or parametric methods \citep{Sersic1963, Sersic1968}. The latter requires explicitly assuming two or more distributions to model the multi-components of galaxies. One of the widely used non-parametric metrics is the asymmetry, concentration, and smoothness system \citep[$ACS$,][]{Conselice2003a}. The asymmetry parameter ($A$) quantifies the lopsidedness and variance in the brightness distribution \citep{Schade1995, Bershady2000, Conselice2000}; the concentration parameter ($C$), which characterizes the light distribution within a galaxy \citep{Abraham1994, Abraham1996}; and the clumpiness parameter ($S$) measures the fragmentation and presence of star-forming regions \citep{Takamiya1999, Conselice2003a}. In addition, other non-parametric morphological indicators such as the Gini$-M_{\rm 20}$, $MID$, $A_{\rm O}-D_{\rm O}$, and $A_{\rm S}$ are also widely employed by researchers to investigate galaxy morphological properties and mergers \citep{Lotz2004, Freeman2013, Wen2014, Pawlik2016}, and understand the fundamental evolutionary processes that shape galaxy formation and evolution \citep[e.g.][]{Conselice2005, Lotz2008a, Whitney2021}.

The impact of varying rest-frame wavelengths on these morphological indicators is significant and has been extensively studied \citep[e.g.][]{Taylor-Mager2007, Baes2020, Yao2023}. \citet{Lotz2004} observed that longer rest-frame wavelengths reveal more extended structures, which they attribute to the predominant presence of older stellar populations. Additionally, \citet{Kelvin2012} highlighted the importance of multi-wavelength morphology in galaxies, spanning from ultraviolet (UV) to near-infrared (NIR) wavelengths. They found that early-type galaxies tend to be more concentrated, redder, and composed of older stellar populations than later-type galaxies. This aligns with the typical redshift dependence, where higher redshift objects appear more compact at shorter wavelengths \citep{Taylor2010}. Therefore, the rest-frame wavelength dependence of morphological indicators emphasizes the need for careful band selection in any morphological analysis. Furthermore, these variations can potentially be leveraged to investigate the stellar mass distribution or stellar population gradients in galaxies \citep{Kelvin2014}. To align morphological indicators at various redshifts to a common rest-frame wavelength in single-band images, the morphological $K$-correction is often employed \citep{Conselice2008, Lopez-Sanjuan2009}. The effects of observational factors like the point spread function (PSF) and noise on morphological parameter measurements are also significant. A low-resolution PSF has a smoothing effect on sharp structural features in observed galaxies, leading to substantial changes in indicators that describe the degree of clumping and light concentration.

Several studies have attempted noise correction in parameter measurements \citep{Shi2009, Wen2016} or used simulated images with varying PSFs and noise levels to account for observed effects \citep{Thorp2021, Yu2023}. When considering {\it JWST}/NICam multi-band data, it's possible to constrain galaxy morphology to an observational effects rest-frame band by selecting different bands at varying redshifts. Yet, it's essential to adjust for the variances in PSFs and noise levels across different filters, calling for the correction of observed morphological indicators. In this study, we leverage the TNG50-simulated galaxies to assess the influence of observational effects on non-parametric morphological measurements, replicating the conditions of the {\it JWST}/NIRCam. We then employ this understanding to correct the non-parametric indicators of galaxies observed by {\it JWST}. Furthermore, we investigate the evolution of galaxies based on color, rest-frame wavelength, redshift, and stellar mass.

The paper is organized as follows: Section 2 describes our data and sample selections. In Section 3, we outline the methodologies used to quantify the effects of observation on non-parametric indicators and correct the indicators characterizing {\it JWST} galaxies. The main findings are outlined in section 4. Subsequently, Section 5 presents a thorough discussion of our results, leading to the final summary and conclusion in Section 6. For our computations, we adopt a concordance cosmology with $H_{0}=$70 km s$^{-1}$ Mpc$^{-1}$, $\Omega_{\rm m}$=0.3, and $\Omega_{\Lambda}$=0.7.

\begin{figure*}[ht]
\centering
\includegraphics[width=1\textwidth] {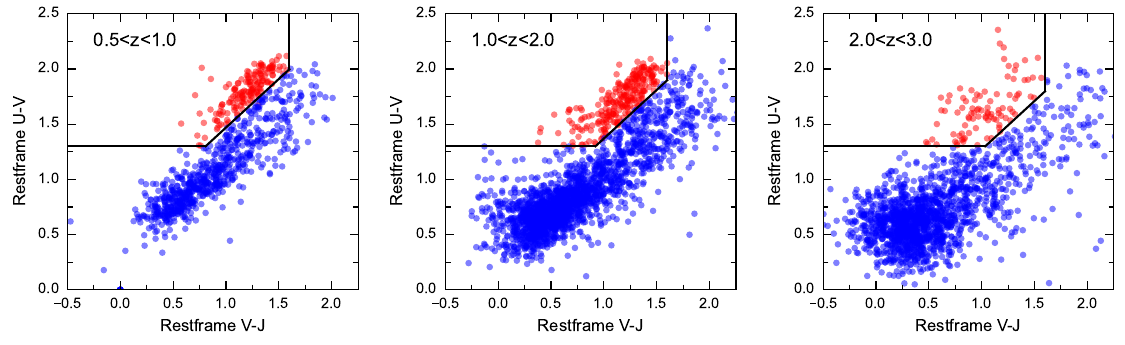} 
\caption{Rest-frame UVJ diagrams at different redshift bins for our sample galaxies selected from the {\it JWST}/CEERS and JADES. The blue and red markers refer to star-forming galaxies (SFGs) and quiescent galaxies (QGs), respectively.}
\label{fig:uvj}
\end{figure*}

\section{Data and Sample } \label{sec:method}

\subsection{The IllustrisTNG Simulation Samples}

The IllustrisTNG project \footnote{\url{https://www.tng-project.org}} \citep{Marinacci2018, Naiman2018, Nelson2018, Nelson2019a, Nelson2019b, Pillepich2018a, Pillepich2018b, Pillepich2019, Springel2018} is a suite of large-volume, cosmological, gravo-magnetohydrodynamical simulations consisting of three sets of simulations encompassing different volumes with varying resolutions. The TNG50, TNG100, and TNG300 simulations cover cubic volumes of 50.17 cMpc, 106.5 cMpc, and 302.6 cMpc on a side, respectively. The IllustrisTNG project includes a comprehensive model for galaxy formation physics, and each TNG simulation self-consistently solves for the coupled evolution of dark matter, cosmic gas, luminous stars, and supermassive black holes from early times to the present day.

The light-cone technique provides a robust approach for establishing a direct connection between theoretical models and observations of galaxies in the distant Universe. It applies to empirical, semi-analytical, and hydrodynamic models, enabling us to trace the evolution of large-scale structures of the Universe and study the morphological characteristics of galaxies. To replicate the observed evolution of galaxies as seen by a hypothetical observer in the present epoch, \citet{Snyder2017} employed the approach developed by \citet{Kitzbichler2007} to convert the 3D positions of galaxies within the IllustrisTNG periodic cube into 3D light-cones. Different geometries, denoted by (m, n) coordinates, were utilized, resulting in the generation of several light-cones: a large 365 arcmin$^2$ light-cone (6,5) based on TNG300-1; a moderately-sized light-cone (7,6) based on TNG100-1; and narrower light-cones (11,10) and (12,11) derived from TNG50-1. These geometries were observed from different vantage points, including the $xyz$, $yxz$, and $zyx$ directions. In total, 12 light-cone catalogs were created, and within each catalog, the stellar mass, redshifts, and other physical parameters were provided.

To construct the light-cone images for the catalogs, \citet{Snyder2023} generated a series of blank square images in FITS format for each light-cone. The central pixel of each image was determined by the right ascension and declination coordinates, which were set at 0,0. The extragalactic field was then populated with IllustrisTNG galaxies positioned accordingly. The pixel size of each image was either 0.03 arcsec (in alignment with {\it HST}/ACS and {\it JWST}/NIRCam), or 0.06 arcsec (for {\it HST}/WFC3), and the flux measurements were provided in nJy. For {\it JWST}/NIRCam, these fields encompassed multiple filters, ranging from F090W to F444W. Additionally, maps of stellar mass and star formation rates (SFRs) were also included \citep[refer][for further details]{Snyder2023}.

We selected three light-cones, TNG50-11-10-xyz, TNG50-11-10-yxz, and TNG50-11-10-zyx, as the targets for our study to investigate the impact of PSF and noise level on galaxy morphology. Each light-cone covers an area of 8 arcmin$^2$ and spans a redshift range from $z=0.1$ to $z=12$ across the three catalogs \citep{Snyder2023}. Subsequently, we selected galaxies with stellar mass $\log(M_*/M_\odot) > 9$ and $0.5 < z < 3$ for our study. We generated $401 \times 401$ cutout images and carefully examined each image to remove any point sources and blended sources. As a result, our final sample consists of over 800 TNG50 galaxies.

\subsection{Observation Data}

\subsubsection{{\it JWST}/ NIRCam imaging data}

The Cosmic Evolution Early Release Science Survey (CEERS; PID 1345, PI Finkelstein) of the James Webb Space Telescope ({\it JWST}) is a comprehensive program designed to demonstrate the capabilities of {\it JWST}, with a specific focus on the study of high-redshift galaxies. The CEERS survey utilizes the Near Infrared Camera (NIRCam) and the Mid-Infrared Instrument (MIRI) to conduct deep imaging covering approximately 100 arcmin$^2$ regions within the CANDELS Extended Groth Strip (EGS) field \citep{Finkelstein2017}. The {\it JWST}/CEERS NIRCam imaging survey comprises 10 pointings, and each pointing was observed using seven filters: F115W, F150W, and F200W on the short-wavelength (SW) side, and F277W, F356W, F410M, and F444W on the long-wavelength (LW) side \citep{Finkelstein2023}. The reduced data from the 10 pointings are now publicly available \citep{Bagley2023}.

The {\it JWST} Advanced Deep Extragalactic Survey (JADES) emphasizes infrared imaging and spectroscopy in the Great Observatories Origins Deep Survey South (GOODS-S) and North (GOODS-N) deep fields to study galaxy evolution. JADES makes use of approximately 770 hours of Cycle 1 observations, primarily utilizing the Near-Infrared Camera (NIRCam) and Near-Infrared Spectrograph (NIRSpec) instruments on {\it JWST}. In the GOODS-S field, focusing on the Hubble Ultra Deep Field and Chandra Deep Field South, JADES achieves deep imaging across an area of approximately 45 arcmin$^2$, with 130 hours of exposure time distributed over nine filters. Moreover, JADES also includes a medium-depth (20 hours) imaging with NIRCam covering approximately 175 arcmin$^2$, using 8-10 filters in both the GOODS-S and GOODS-N fields \citep{Eisenstein2023}. The JADES first data release (DR1) and second data release (DR2) cover an area of $\sim$ 80 arcmin$^2$ in the GOODS-S field, and {\it JWST}/NIRCam imaging data and photometric catalog are now publicly available \citep{Rieke2023, Hainline2023}.

\subsubsection{Multiwavelength data}

The Cosmic Assembly Near-infrared Deep Extragalactic Legacy Survey \citep[CANDELS,][]{Grogin2011, Koekemoer2011} contains five extragalactic deep fields with {\it HST}, {\it Spitzer}, and ground-based imaging data from UV to 24\,$\mu$m. Using CANDELS multiwavelength data, \citet{Stefanon2017} estimated the photometric redshift, stellar mass, and other physical parameters of the galaxies in the CANDELS Extended Groth Strip (EGS) field. \citet{Santini2015} combined the stellar masses using the same photometry and redshifts from 10 different teams to obtain the median stellar mass of the GOODS-S and UDS fields. Hathi (N. Hathi, private communication) collected spectroscopic redshifts and grism redshifts in the CANDELS fields. We additionally collected the MUSE-wide survey data \citep{Urrutia2019}, MUSE Hubble Ultra-Deep Field survey data \citep{Bacon2023} in the GOODS-S field.

\subsubsection{Sample}
We cross-match the CANDELS, median stellar mass catalog, and redshifts catalog using an aperture with a radius of 0.54 arcsec. Then, we select sources with $\log(M_*/M_\odot) > 9$ and $0.5 < z < 3$ as the parent sample in the {\it JWST}/CEERS covered EGS field and {\it JWST}/JADES covered GOODS-S field. The {\it JWST}/NIRCam images of each parent sample are PSF-matched to F444W. We stack 5 PSF-matched images and F444W images to perform segmentation for each source. To ensure the detection of galaxies in all filters and remove any point sources, we carefully inspected the images and segmentation maps obtained from the 6 NIRCam filters (F115W, F150W, F200W, F277W, F356W, and F444W). In cases where segmentation issues were identified, we manually adjusted the parameters of the \texttt{photutils} package and re-did the image segmentation to improve the accuracy of morphological measurements. As a result, we obtained a sample of over 4600 galaxies with relatively reliable rest-frame parameters and {\it JWST}/NIRCam images. We divide our sample into star-forming galaxies and quiescent galaxies based on the rest-frame $UVJ$ diagrams proposed by \citet{Williams2009}, to study the differences in the evolution of the morphology of these two classes of galaxies with rest-frame wavelength, redshift, and stellar mass, respectively.

\subsubsection{Rest-frame Band Selection}

\begin{figure}[!t]
\centering
\includegraphics[width=1\columnwidth] {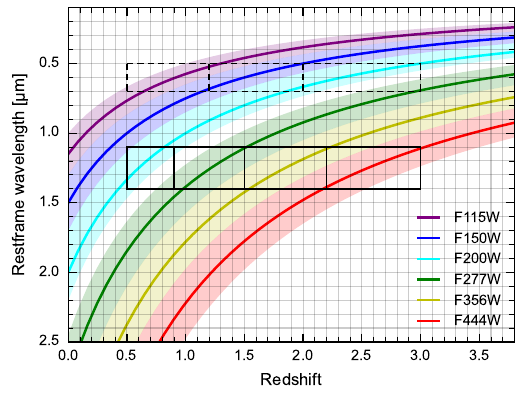} 
\caption{The rest-frame wavelength range of different {\it JWST}/NIRCam filters as a function of redshift is presented. The curves represent the central wavelength of each filter, and the shaded regions represent the bandwidth of each filter. Dashed and solid boxes indicate the selection ranges of the rest-frame V-band and J-band, respectively.}
\label{fig:z_wavelength}
\end{figure}

We chose two fixed rest-frame bands to investigate the galaxy morphology evolution with redshift and stellar mass. Specifically, we selected two wavelength ranges, 0.5-0.7$\,$ $\mu$m, and 1.1-1.4$\,$ $\mu$m, based on Figure \ref{fig:z_wavelength}. We refer to these bands as the rest-frame V-band and rest-frame J-band, respectively. The selection of specific filters within each band is based on the redshifts of the sample galaxies: F115W for $0.5 < z < 1.2$ in the V-band, F150W for $1.2 < z < 2.0$, and F200W for $2.0 < z < 3.0$. For the rest-frame J-band, we used F200W for $0.5 < z < 0.9$, F277W for $0.9 < z < 1.5$, F356W for $1.5 < z < 2.2$, and F444W for $2.2 < z < 3.0$.

\begin{figure}[!t]
\centering
\includegraphics[width=1\columnwidth] {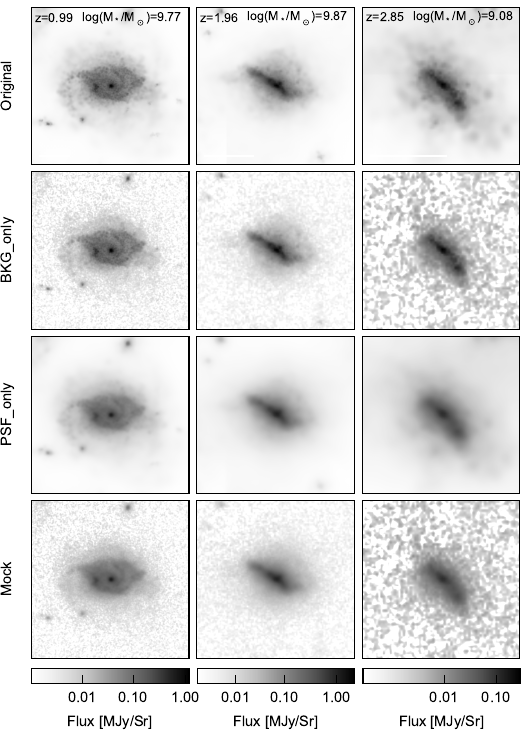} 
\caption{ From top to bottom panels: noise-free images (\texttt{Original}), images with only noise added to the original images (\texttt{Noise-only}), images with only the PSF added (\texttt{PSF-only}), and images with both noise and the PSF added (\texttt{Mock}) for three galaxies at varying redshifts.}
\label{fig:Mock_images}
\end{figure}

\section{Methodology}

\subsection{Observational Effects on TNG50 Galaxies}

First, we identify over 20 isolated stars from the CEERS field and stack all stars in each band to create empirical PSF models. Additionally, we carefully select several background regions in the CEERS field to estimate the noise level in each band and generate a noise map for the $401 \times 401$ pixel images across the six {\it JWST}/NIRCam broad bands (F115W, F150W, F200W, F277W, F356W, F444W). We generate various types of images for each simulated image using these components.

\begin{enumerate}
\item The \texttt{Original} images, obtained directly from the TNG50 mock survey \citep{Snyder2023}, are considered to be free from any observational effects.

\item The \texttt{Noise-only} images are created by overlaying the \texttt{Original} images with the corresponding noise maps.

\item The \texttt{PSF-only} images are produced by convolving the \texttt{Original} images with the empirical PSF in each band.

\item The \texttt{Mock} images are formed by overlaying the noise maps on the \texttt{PSF-only} images obtained in step 3. These \texttt{Mock} images imitate the TNG50 simulated galaxies observed under the {\it JWST} observational conditions.
\end{enumerate}

Analogous to the {\it JWST} observed galaxy images, we conduct segmentation on the \texttt{Mock} images using the PSF-matched to F444W stacked images. Figure \ref{fig:Mock_images} presents examples of the four types of simulated images mentioned above.

\subsection{The Morphological indicator measurements}

\begin{figure*}[!t]
\centering
\includegraphics[width=1\textwidth] {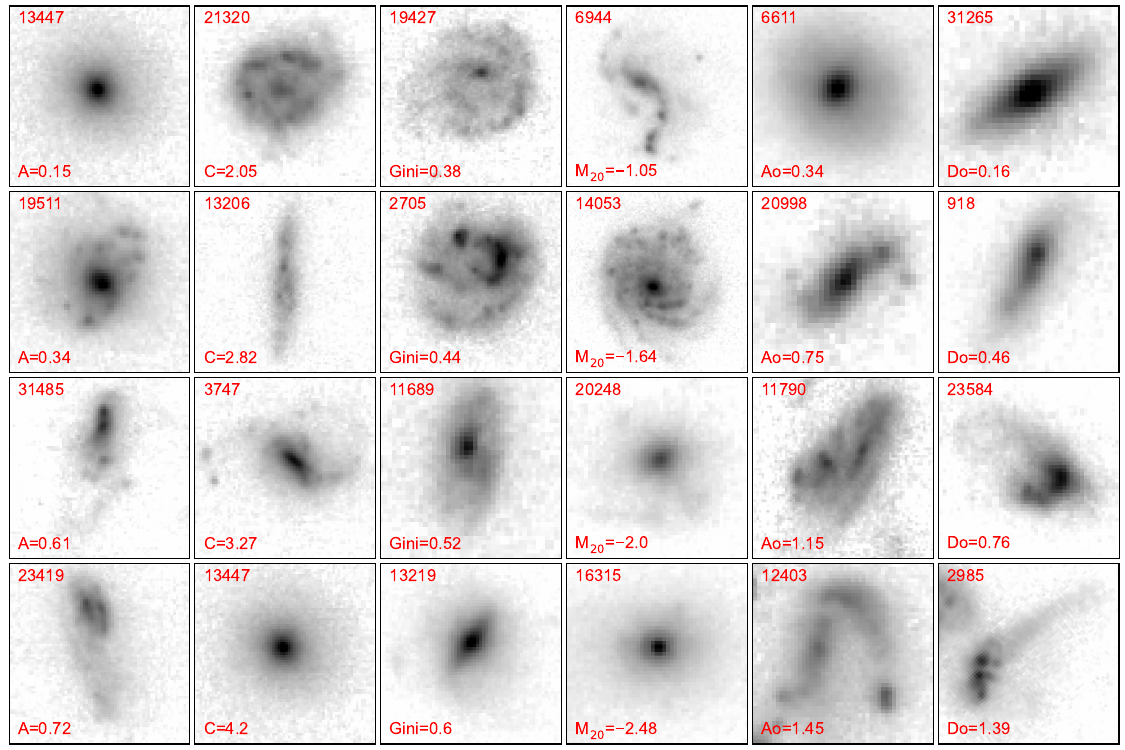} 
\caption{Example images showing variations in morphological indicators obtained in the F200W band from {\it JWST}/CEERS are presented. The IDs of objects from the CANDELS/EGS photometry catalog are displayed in the left-top corner. The measured indicators are displayed in the left-bottom corner. }
\label{fig:Example_images}
\end{figure*}

We employ the code developed by \citet{Ren2023} to measure these commonly used non-parametric indicators and the signal-to-noise ratio (S/N) for the TNG50 galaxies and {\it JWST} observed galaxies. For the TNG50 galaxies, the same segmentation is applied to measure the non-parametric morphological indicators ($C$, $A$, $S$, $Gini$, $M_{20}$, $A_{\rm O}$, $D_{\rm O}$) and the Petrosian radius ($r_{\rm p}$) in all six bands for the \texttt{Original}, \texttt{Noise-only}, \texttt{PSF-only}, and \texttt{Mock} images. Some examples of {\it JWST}/CEERS F200W images are shown in Figure \ref{fig:Example_images}. We provide a detailed description of the calculation methods for each of these indicators.

\subsubsection{Petrosian radius}

The vast majority of morphological indicator measurements rely on the Petrosian radius \citep[$r_{\rm p}$,][]{Petrosian1976}, which is defined as the radius at which the surface brightness is equal to a certain fraction $\eta$ of the mean surface brightness within $r_{\rm p}$:
\begin{equation}
\eta(r) = \frac{I(r)}{\langle I(r) \rangle},
\end{equation}
where $\eta(r)$ is the Petrosian index at $r$. $I(r)$ represents the surface brightness, and $\langle I(r) \rangle$ represents the mean surface brightness within $r$. We use $\eta(r_{\rm p}) = 0.2$ for our measurements.

\subsubsection{$ACS $ indicators}

The asymmetry parameter ($A$) measures the rotational asymmetry of a galaxy image compared to the asymmetry of its background noise. Previous studies have noted that high levels of noise can lead to an overestimation of background asymmetry, resulting in a smaller $A$ parameter in the final calculation. To address this issue, we utilize a noise-corrected asymmetry algorithm introduced by \citet{Wen2016}:
\begin{equation}
A = \frac{\sum{|I_{\rm 0} - I_{\rm 180}|} - \delta_2}{\sum{|I_{\rm 0}|} - \delta_1},
\end{equation}
where $\delta_1 = f_1 \times \Sigma |B_{\rm 0}|$, $f_1 = N_{\rm flux < 1 \sigma} / N_{\rm all}$, $\delta_2 = f_2 \times \Sigma |B_{\rm 0} - B_{\rm 180}|$, and $f_2 = N_{|\rm flux| < \sqrt{2}} / N'_{\rm all}$. Here, $I_{\rm 0}$ and $I_{\rm 180}$ refer to the original high-resolution image and its 180$^\circ$ rotated counterpart. Similarly, $B_{0}$ represents a background patch with the same shape as $I_{\rm 0}$, and $B_{180}$ is the 180$^\circ$ rotation of $B_{0}$. The correction factors, $\delta_1$ and $\delta_2$, account for noise contributions to the flux image $I_{\rm 0}$ and the residual image $I_{\rm 180}$, respectively. The rotational center ($x_{\rm a}$, $y_{\rm a}$) is defined as the point where $I_{\rm 0} - I_{\rm 180}$ is minimized, and all operations are restricted to an elliptical aperture centered at ($x_{\rm a}$, $y_{\rm a}$) with the major axis of 1.5 $r_{\rm p}$.

The light concentration parameter ($C$) measures the extent to which the light in a galaxy image is concentrated towards its center \citep[e.g.][]{Abraham1994, Bershady2000, Conselice2003a}. It is commonly calculated as:
\begin{equation}
C = 5 \times \text{log}(\frac{r_{80}}{r_{20}}),
\end{equation}
where $r_{80}$ and $r_{20}$ represent the radii of circular apertures enclosing 80\% and 20\% of the galaxy's total flux, respectively. The rotation center serves as the center for these circular apertures.

The smoothness parameter ($S$) describes the structural characteristics of galaxies, such as clusters, spiral arms, and other substructures. We adopt the formula proposed by \citep{Conselice2003a}:
\begin{equation}
S = 10 \times \left[\left(\frac{\sum(|I_{0} - I_{0}^{\sigma}|)}{\Sigma |I_{0}|}\right) - \left(\frac{\sum(|B_{0} - B_{0}^{\sigma}|)}{\sum |I_{0}|}\right)\right],
\end{equation}
where $I_{0}$ and $B_{0}$ are the original image and background image, while $I_{0}^{\sigma}$ and $B_{0}^{\sigma}$ represent the smoothed image and smoothed background, respectively. The smoothing is performed using a boxcar kernel with a size of $\sigma = 0.25 \times r_{\rm p}$. The sum is carried out over all pixels at distances between 0.25 $r_{\rm p}$ and 1.5 $r_{\rm p}$ from the rotation center \citep{Lotz2004}.

\subsubsection{$G-M_{\rm 20}$ indicators}

The $G$ parameter and the $M_{\rm 20}$ parameter are measured for pixels within the Gini-segmentation map of galaxy images. We use the method given by \citet{Lotz2004} to produce a Gini-segmentation map. We first smooth the galaxy image using a Gaussian kernel with $\sigma = 0.2 r_{\rm p}$, and then obtain the Gini-segmentation map by setting the value of the pixel with a flux above the mean flux at $r_{\rm p}$ and below 10 $\sigma$ to 1, and other pixels to 0.

The Gini coefficient ($G$) is used to study the distribution of light on each pixel in galaxy images \citep{Lotz2004}. It can be computed as,
\begin{equation}
G = \frac{1}{|\bar{f}|n(n-1)} \sum_{i}^{n} (2i - n - 1) |f_{i}|,
\end{equation}
where $\bar{f}$ refers to the average flux per pixel, $n$ is the number of pixels within the Gini-segmentation map, $i$ ranges from 0 to $n$, and $f_{i}$ is the flux of the $i$-th pixel. The $G$ takes a value between 0 and 1, with $G = 0$ indicating that the flux in all pixels is the same ($|\bar{f}|$), and $G = 1$ indicating that all flux within the Gini segmentation is concentrated in one pixel.

The second-order moment parameter ($M_{\rm 20}$) can be calculated as the ratio between the second-order moments of the pixels where the brightest 20\% of the light is located and the second-order moments of all the pixels within the Gini segmentation. The formula provided by \citet{Lotz2004} is as follows:

\begin{equation}
M_{20} = {\rm log}\left(\frac{\sum_{i} M_{i}}{M_{\rm tot}}\right), \ {\rm while} \ \sum f_{i} < 0.2 \times f_{\rm tot},
\end{equation}
\begin{equation}
M_{\rm tot} = \sum_{i}^{n} M_{i} = \sum_{i}^{n} f_{i}[(x_{i} - x_{\rm m})^2 + (y_{i} - y_{\rm m})^2],
\end{equation}
where $f_{i}$ is the flux value of the $i$-th pixel from the largest to the smallest within the Gini-segmentation map. $(x_{\rm m}, y_{\rm m})$ is the moment center pixel, which minimizes the $M_{\rm tot}$.

\subsubsection{$A_{\rm O}-D_{\rm O}$ indicators}

\citet{Wen2014} introduced the $A_{\rm O}-D_{\rm O}$ for detecting asymmetric structures in the galaxy outskirts. This method involves dividing the galaxy images into the outer half-light region (OHR) and the inner half-light region (IHR). The two indicators are obtained by computing the OHR asymmetry $A_{\rm O}$ and the relative deviation of the IHR center and OHR center $D_{\rm O}$. Our calculations in this paper do not directly subtract a circular aperture containing 50\% of the light centered at ($x_{\rm a}$, $y_{\rm a}$) as is the case for \texttt{STATMORPH} \citep{Rodriguez-Gomez2019}. We utilize the approach of \citet{Wen2016} to define OHR and IHR. First, all the pixels in the galaxy image are arranged in order of their flux, from brightest to faintest. We begin the selection of pixels from the brightest end of this arrangement, defining $f$ as the ratio of the total flux of the selected pixels to the total flux of the galaxy. As we gradually increase $f$, independent pixel groups tend to form from the brightest selected pixels. We start by selecting pixels accounting for half of the total flux of the galaxy ($f=50\%$), and these pixels tend to form one or several independent pixel groups in images. We calculate the flux of each pixel group and continue to increase $f$ until the flux of the brightest pixel group reaches 25\% of the total flux of the galaxy. We then calculate the centroid of the brightest pixel group and use it as the center to fit an ellipse to the pixel group. We fix the axis ratio of this ellipse and gradually increase its major axis. When the flux within the elliptical aperture reaches 50\% of the total flux of the galaxy, the ellipse is used to divide the galaxy image into IHR and OHR.

The outer asymmetry $A_{\rm O}$ is defined as follows,
\begin{equation}
A_{\rm O} = \frac{\sum{|I_{\rm 0} - I_{\rm 180}|} - \delta_2}{\sum{|I_{\rm 0}|} - \delta_1},
\end{equation}
where $\delta_1 = f_1 \times \Sigma |B_{\rm 0}|$. $f_1 = N_{\rm flux < 1 \sigma} / N_{\rm all}$, $\delta_2 = f_2 \times \Sigma |B_{\rm 0} - B_{\rm 180}|$, and $f_2 = N_{|\rm flux| < \sqrt{2}} / N'_{\rm all}$. $I_{\rm 0}$ and $I_{\rm 180}$ refer to the OHR image and 180$^{\circ}$-rotated OHR image, respectively. Similarly, $B_{0}$ is a background patch in the image with the same shape as $I_{\rm 0}$. $B_{180}$ is the 180$^\circ$ rotation of $B_{0}$. The two correction factors, $\delta_{1}$ and $\delta_{2}$, are noise contributions to the flux image $I_{\rm 0}$ and the residual image $I_{\rm 180}$, respectively. The number fraction of pixels in the OHR that are dominated by noise is represented by $f_{1}$. $f_{2}$ represents the number fraction of OHR pixels that are dominated by noise in the residual image. The total number of pixels in the OHR and residual is represented by $N_{\rm all}$ and $N'_{\rm all}$, respectively. The standard deviation of noise in $I_{\rm 0}$ is represented by $\sigma$. The centroid of the whole galaxy is used as the rotational center of the OHR. See \citet{Wen2014, Wen2016} for more details.

\begin{figure}[!h]
\centering
\includegraphics[width=1\columnwidth] {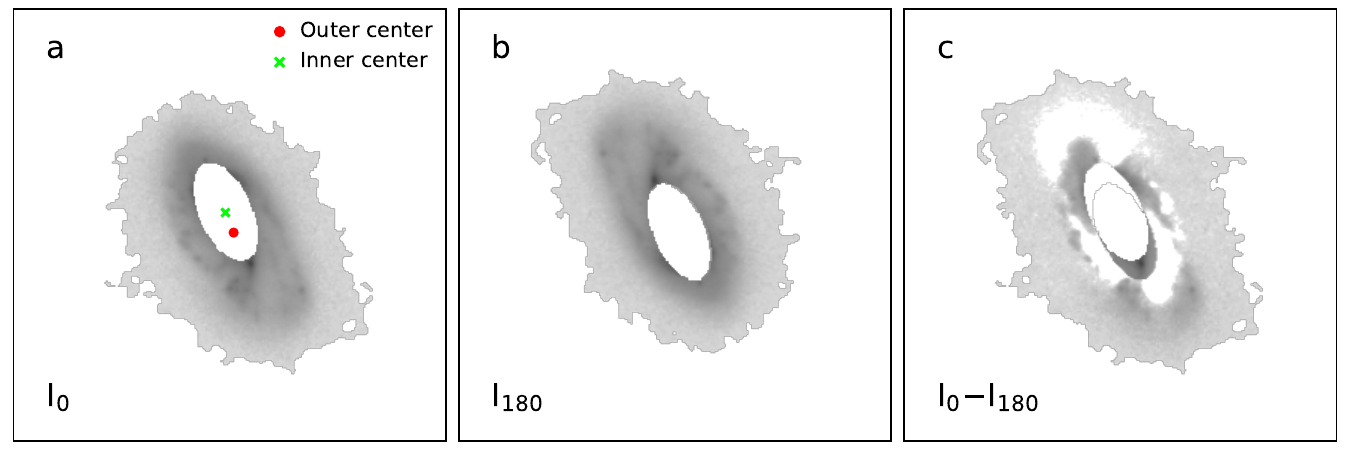} 
\caption{An example of the $A_{\rm O}$ and $D_{\rm O}$ measurement is as follows. ${\bf a,}$ the outer half-light region (OHR) image of the F200W band in the JWST/CEERS; ${\bf b,}$ the 180$^{\circ}$-rotated image of the OHR; ${\bf c,}$ the residual image of the OHR.}
\label{fig:ao-do}
\end{figure}

The outer deviation $D_{\rm O}$ is defined as follows,
\begin{equation}
D_{\rm O} = \frac{\sqrt{(x_{\rm O} - x_{\rm I})^2 + (y_{\rm O} - y_{\rm I})^2}}{r_{\rm eff}},
\end{equation}
where ($x_{\rm I},$ $y_{\rm I}$) and ($x_{\rm O},$ $y_{\rm O}$) refer to the centroids of the IHR and OHR, respectively. The $r_{\rm eff}$ is defined as $\sqrt{(n / \pi)}$, where $n$ is the pixel number of the IHR. Figure \ref{fig:ao-do} shows the OHR image, 180$^{\circ}$-rotated OHR image, and residual images of a galaxy.

\subsection{Corrected Mock Morphological  Indicators to Original Morphological Indicators}

\begin{figure*}[!ht]
\centering
\includegraphics[width=1\textwidth] {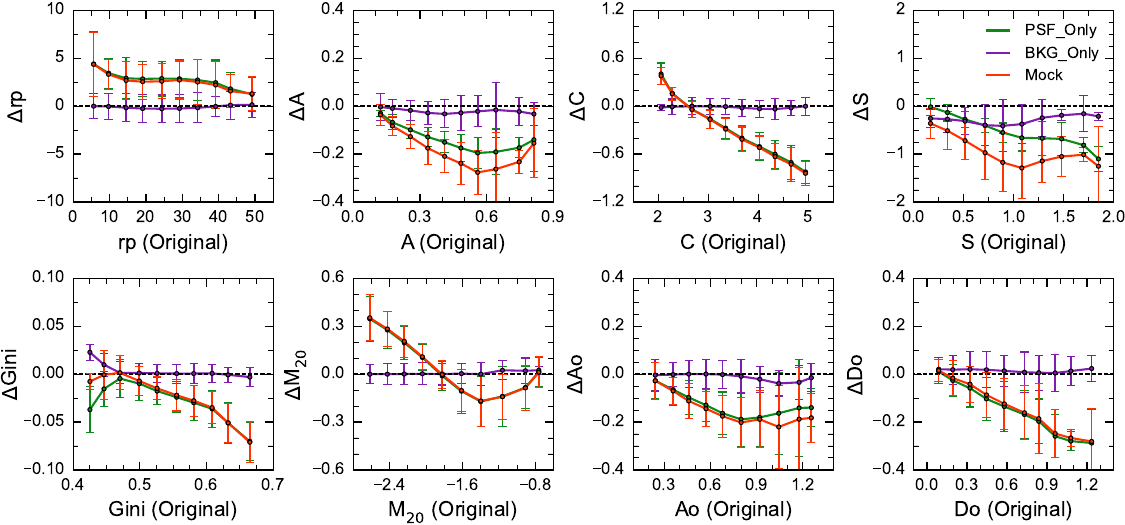} 
\caption{The deviations of non-parametric morphological indicators as a function of the true indicators in the F200W filter, after accounting for observational effects added to TNG50 simulated images.}
\label{fig:Non-para_deviations}
\end{figure*}

We computed the discrepancies in the galaxy morphological indicators between the \texttt{Original} images and \texttt{Noise-only}, \texttt{PSF-only}, and \texttt{Mock} images across all six filters. The variations in morphological indicators resulting from the three different effects in the F200W band are illustrated in Figure \ref{fig:Non-para_deviations}. Our findings reveal that the non-parametric indicators of the TNG50 galaxies in our sample are minimally affected by the {\it JWST}/CEERS noise level when the PSF is absent. However, there are some impacts on noise-sensitive indicators, such as $S$. Under the \texttt{PSF-only} conditions, the morphological indicators of galaxies exhibit significant deviations from the \texttt{Original} images, with this effect varying according to the \texttt{Original} indicators. When both the PSF and noise are present, the measured values of morphological indicators deviate noticeably from their true values.

To mitigate this issue, we apply a correction to the indicators derived from the \texttt{Mock} images, aiming to align them more closely with the indicators measured without observational effects. For each parameter in each band, we implement the following corrections (using the $A$ parameter as an example):
\begin{equation}
A = \alpha \cdot A_{\rm obs}^2 + \beta \cdot A_{\rm obs} + \gamma,
\end{equation}
where $A$ represents the corrected indicators and $A_{\rm obs}$ denotes the observed indicators, which are measured in the \texttt{Mock} images for the TNG50 sample. Given that it is not feasible to directly determine the morphological indicators of galaxies in the absence of observational effects, this correction factor can only be inferred from observations. Through numerous iterations, we determined that a second-order polynomial provides a better fit for the observed (\texttt{Mock}) indicators and the \texttt{Original} indicators.

For each filter, specific ($\alpha$, $\beta$, $\gamma$) values associated with each morphological indicator are provided in Appendix \ref{Correction index}. However, the analysis excludes the $S$ parameter due to its sensitivity to the PSF and noise. The deviations observed in the different morphological indicators are correlated with redshift. This can be attributed to the varying effects of the PSF caused by the diverse sizes of galaxies at different redshifts, as well as the evolution of morphological indicators with redshift and mass. Further detailed discussion on this topic is presented in \ref{Primary Factors Influencing the Morphological Parameters}.

\begin{figure*}[!ht]
\centering
\includegraphics[width=1\textwidth] {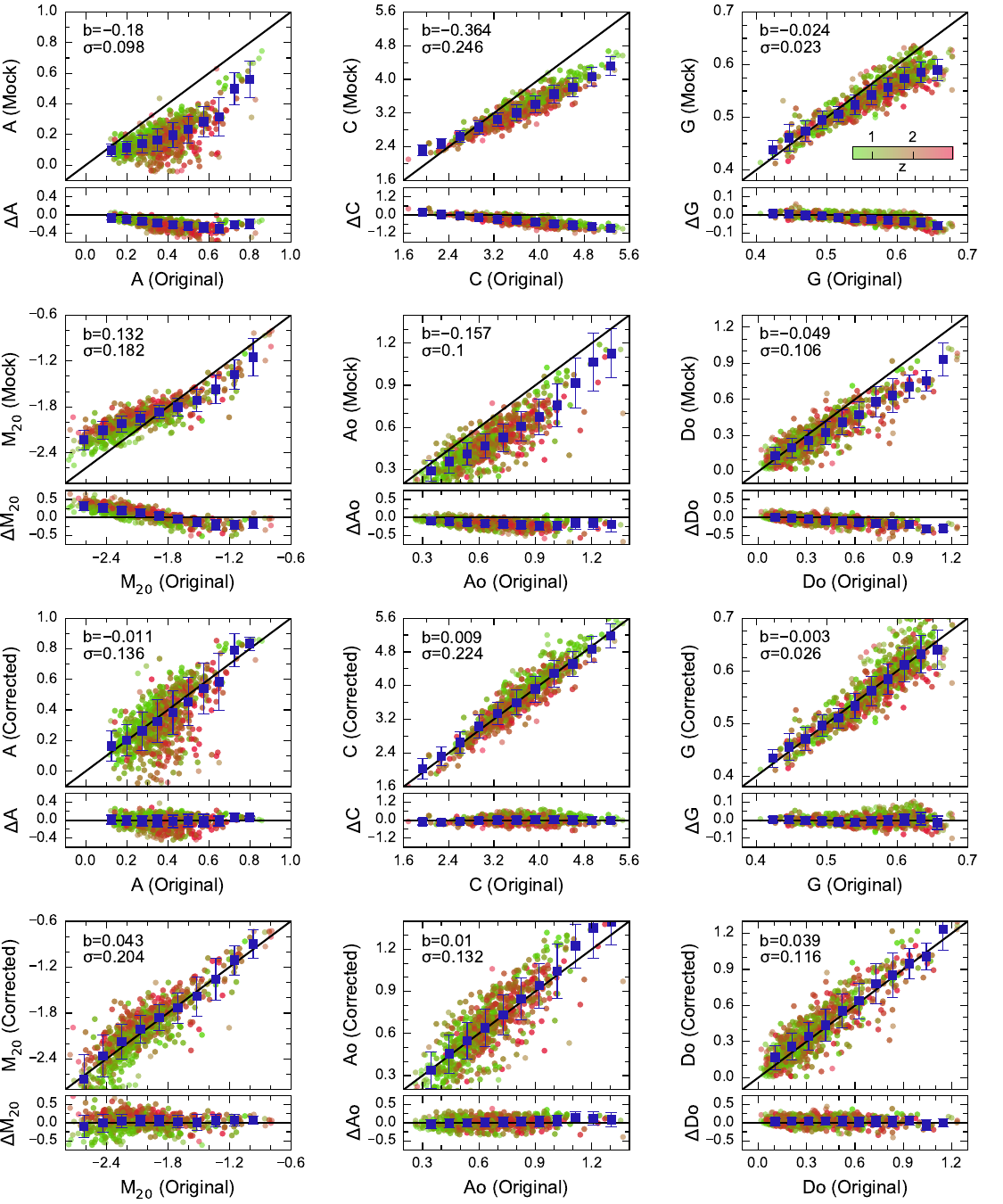} 
\caption{The top six panels show the comparisons between the Original and Mock indicators of TNG50 galaxies in the {\it JWST}/NIRCam F200W filter, while the bottom six panels show the corrected indicators. The correction functions can be found in the text. The error bars represent the scatter within 68\% of the samples in each bin.}
\label{fig:corrected}
\end{figure*}

Figure \ref{fig:corrected} illustrates the disparities between the indicators before and after corrections using the \texttt{Mock} images, as well as those without any observational effects. The corrected indicators effectively eliminate the systematic bias introduced by factors such as the PSF and noise. Notably, the fits are better for indicators such as $C$, $A_{\rm O}$, $D_{\rm O}$, and $Gini$. However, some residual bias remains for the $A$ parameter. This can be ascribed to the measurement method of this parameter and the differential impact of noise on various parameter values. The corrections slightly increase the scatter of all indicators. Therefore, we caution that our corrections are mainly suitable for rectifying systematic biases in large-sample analyses and are not accurate enough to correct individual galaxies or infer statistical trends for small samples.

Lastly, we apply the corrections to the observed galaxy images from {\it JWST}, where each indicator is adjusted using its respective ($\alpha$, $\beta$, $\gamma$) polynomial coefficients for each filter. The corrected indicators obtained through this methodology are subsequently utilized for the rest of the paper.

%%%%%%%%%%%%%%%%%%%%%%%%%%%%%%%%%%%%%%%%%%%%%%%%%%%%%%%%%%%%%%%%%%%%%%%%%%%%%%%%%%%%%%%%%%%%%%%%%%%
\section{RESULTS}

\begin{figure*}[!t]
\centering
\includegraphics[width=1\textwidth] {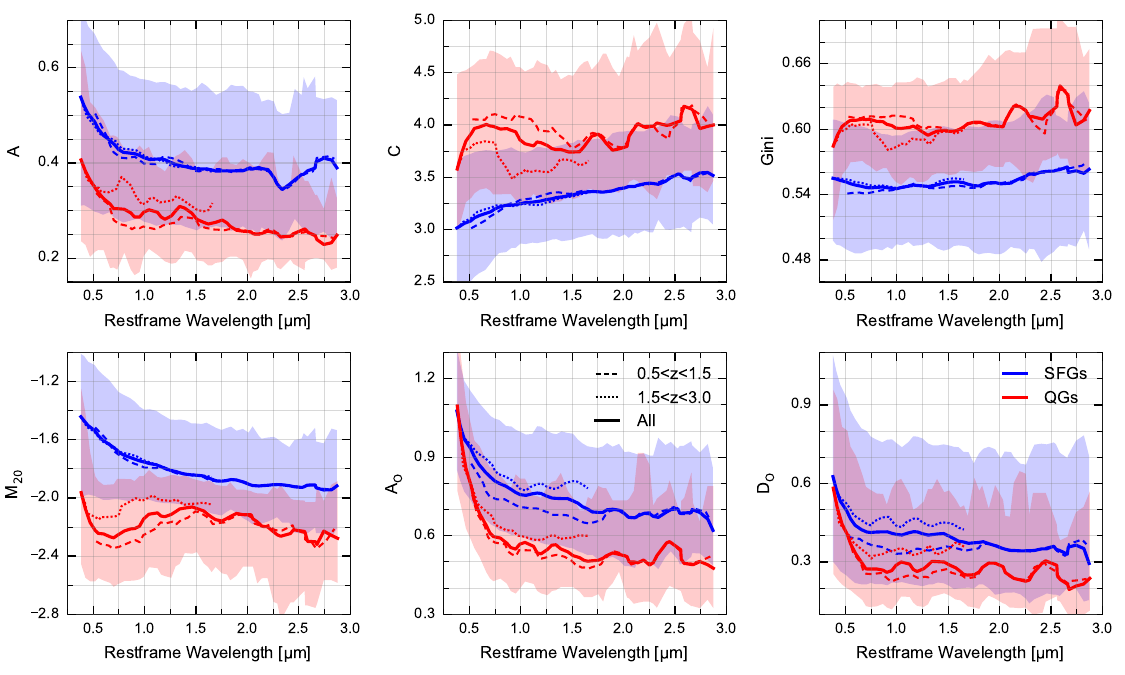} 
\caption{The wavelength dependence of the corrected non-parametric morphological indicators of the {\it JWST}/CEERS and JADES galaxies is shown. The blue solid and red solid lines refer to the star-forming galaxies (SFGs) and quiescent galaxies (QGs) of all samples, respectively. The filled regions represent the range containing 16\% to 84\% of all samples (solid line). The dashed and dotted lines represent the low redshift ($0.5 < z < 1.5$) and high redshift ($1.5 < z < 3$) subsamples, respectively.}
\label{fig:wavelength_dependence}
\end{figure*}

\subsection{Wavelength Dependence of Non-parametric Morphology indictators}

In this analysis, we divided the SFGs and QGs into two redshift ranges, $0.5 < z < 1.5$ and $1.5 < z < 3$, to study the evolution of non-parametric morphological indicators, which is not well understood. Each subsample has corrected morphological indicators for the six {\it JWST}/NIRCam filters. For the galaxies in each subsample, we used the central wavelength $\lambda$/($1+z$) of each filter to represent the rest-frame wavelength ($\lambda_{\rm rf}$). We then divided each subsample into several bins according to the rest-frame wavelength and took the median value of the morphological indicators in each bin. Our results are shown in Figure \ref{fig:wavelength_dependence}. Our results show that the morphological indicators depend on wavelength, especially in the optical bands, as expected. There is a pronounced evolution at $\lambda_{\rm rf} <$ 1\,$\mu$m, with the morphological indicators exhibiting more significant variations with wavelength as $\lambda_{\rm rf}$ becomes shorter. Above approximately $\lambda_{\rm rf} >$ 1\,$\mu$m, the morphological indicators show only slight variation with rest-frame wavelength. On the one hand, indicators that describe the distribution of light within galaxies, such as $C$, $Gini$, and $M_{\rm 20}$, display opposite trends at shorter wavelengths in SFGs and QGs. This difference might arise from variations in the distribution of star formation activity between SFGs and QGs. SFGs have star formation activity primarily distributed in star clusters and spiral arms, contributing more short-wavelength luminosity to the extended disk of the galaxy. On the other hand, the indicators $A$, $A_{\rm O}$, and $D_{\rm O}$, which characterize the flux distribution of galaxies in two dimensions, are primarily related to the nature of their stellar dynamics and star formation rates (SFRs). These indicators exhibit similar trends with morphology and wavelength in both SFGs and QGs galaxies.

\subsection{Morphological Evolution}

\subsubsection{Rest-frame V-band morphology}
\label{V-band}

\begin{figure}[!h]
\centering
\includegraphics[width=1\columnwidth] {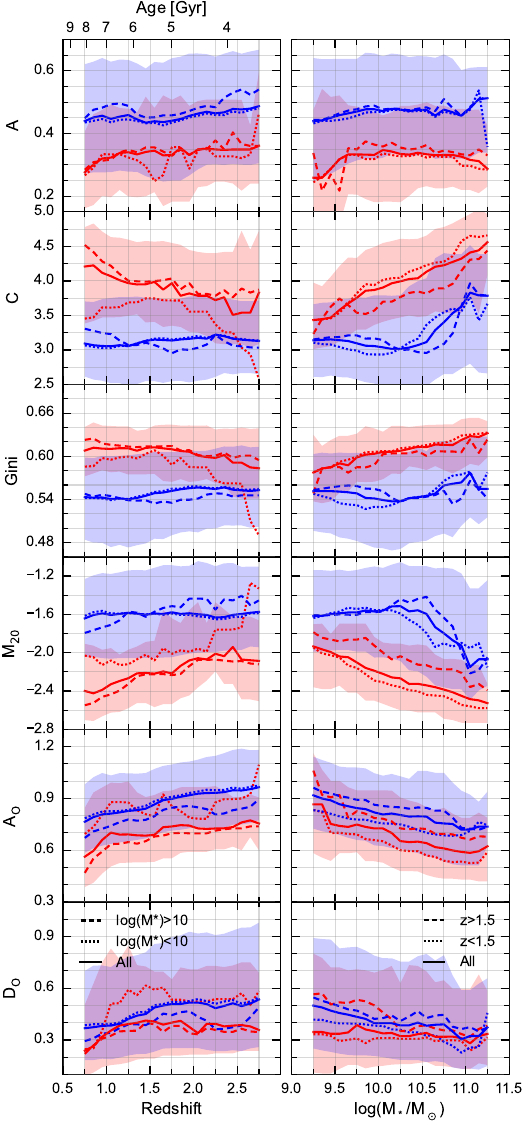} 
\caption{The redshift and stellar mass dependence of the corrected rest-frame V-band non-parametric indicators are presented. The non-parametric indicators are displayed as a function of redshifts in the left panels and as a function of stellar mass in the right panels. The quiescent galaxies (QGs) and star-forming galaxies (SFGs) selected using the UVJ diagrams are represented by the red and blue lines, respectively.}
\label{fig:Rest_V_evo}
\end{figure}

\begin{figure}[!h]
\centering
\includegraphics[width=1\columnwidth] {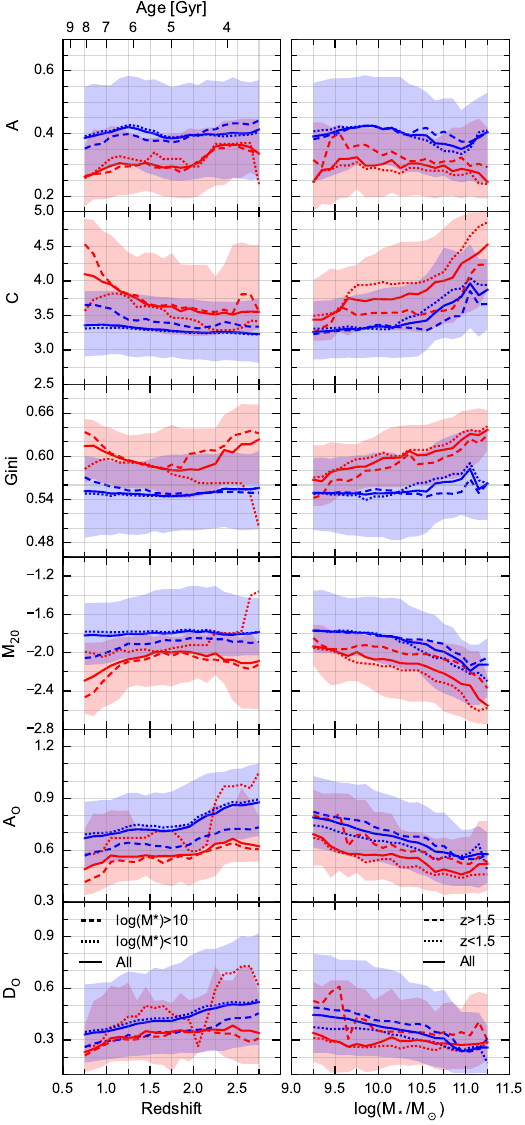} 
\caption{The corrected rest-frame J-band non-parametric indicators are plotted as a function of redshift and stellar mass. All markers are consistent with those used in Figure \ref{fig:Rest_V_evo}.}
\label{fig:Rest_J_evo}
\end{figure}

To examine the relationships between various morphological indicators and redshift/stellar mass, we calculate the median value of each parameter within each window, as well as the scatter in the range of 16\%-84\%, using a sliding window of width $\delta z = 0.5$ for all samples with a signal-to-noise ratio ($S/N$) greater than 2.5 in the rest-frame V-band. The same method is applied to estimate the morphological indicators as a function of stellar mass. The results are shown in Figure \ref{fig:Rest_V_evo}.

As illustrated in the left panels of Figure \ref{fig:Rest_V_evo}, only the $A_{\rm O}$ and $D_{\rm O}$ indicators exhibit an evolutionary trend with redshift for SFGs. This indicates that the morphology of SFGs gradually transitions from externally elongated structures or lopsided structures at high redshifts to a more rounded and Hubble-type appearance. On the other hand, the morphological indicators of QGs show substantial variations with redshift at $z = 1.5$, but this trend gradually diminishes at higher redshifts. These findings suggest that before $z \sim 2$, the morphology of QGs undergoes minimal changes compared to that at lower redshifts. Additionally, the differences between SFGs and QGs are less pronounced around $z \sim 3$ but become more noticeable as the redshift decreases. This indicates that the morphology-galaxy property relations at low redshifts may not be directly applicable to medium to high redshifts.

The right panels of Figure \ref{fig:Rest_V_evo} show the evolution of the morphological indicators with stellar mass. The morphologies of both SFGs and QGs undergo significant changes as the stellar mass increases. Except for the asymmetry parameter $A$, the morphology of QGs exhibits a monotonically increasing trend with increasing galaxy stellar mass. Conversely, for SFGs, indicators related to the concentration of light towards the center ($C$, $Gini$, $M_{\rm 20}$) remain relatively stable until a mass threshold of $\log(M_*/M_\odot) = 10.5$, after which they increase with stellar mass. This shift signifies a redistribution of light towards the galaxy's center and a decrease in light distributed in outer structures such as star clusters in the disk of galaxies. This mass threshold may be linked to the growth of galaxy stellar haloes and the suppression of star formation in galaxies when the dark matter haloes reach a certain mass. Moreover, the differences in morphology between SFGs and QGs in the optical band are more dominant at higher stellar mass ranges than at lower stellar mass ranges.

\subsubsection{Rest-frame J-band morphology}

Likewise, we present the relationship among morphological indicators, redshift, and stellar mass in Figure \ref{fig:Rest_J_evo}. Our results demonstrate that the rest-frame J-band morphological evolution relations are similar to those in the rest-frame V-band but are not as pronounced as the relations in the V-band. Meanwhile, in the J-band, SFGs and QGs exhibit similar evolutionary trends, and the scatter of most morphological indicators in the rest-frame J-band is 5\% to 30\% smaller than that in the rest-frame V band for star-forming galaxies (SFGs), and 5\% to 55\% for quiescent galaxies (QGs). This is primarily because the rest-frame V-band is sensitive to the star formation rates (SFRs) and dust distributions of galaxies. The differences in these physical parameters between SFGs and QGs lead to more considerable morphological differences. On the other hand, the J-band morphology is mainly influenced by older stellar populations. Morphological indicator differences are smaller for galaxies of the same redshift and mass. Additionally, the dispersions of the morphological indicators for both SFGs and QGs are also smaller in the J-band than in the V-band, suggesting that the morphological differences between galaxies are less pronounced in the J-band.

\subsubsection{The morphological evolution of SFGs: accompanying stellar mass growth }

\begin{figure*}[t!]
\centering
\includegraphics[width=0.99\textwidth] {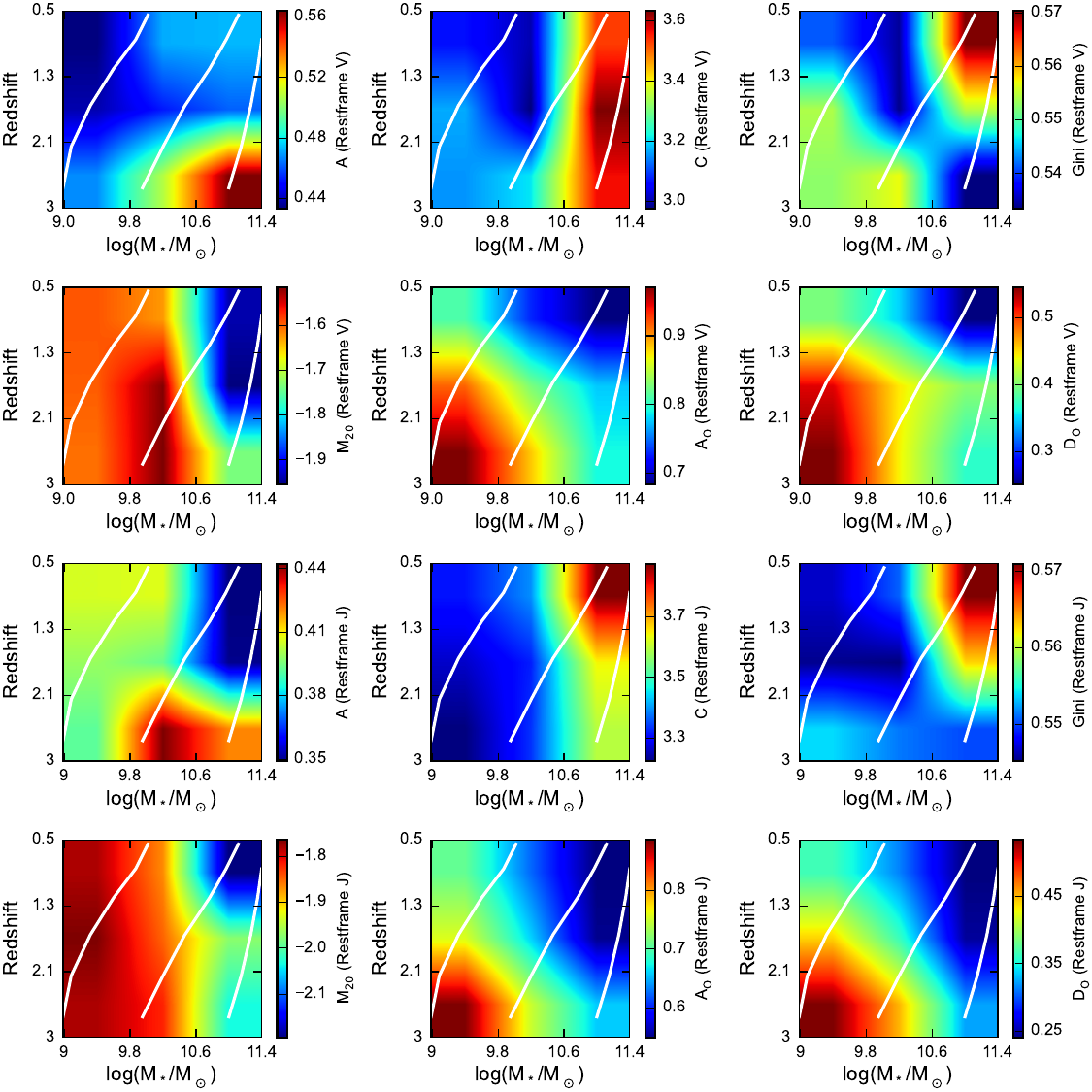} 
\caption{The distributions of median morphological indicators in the rest-frame V-band (top three panels) and rest-frame J-band (bottom three panels) in the redshift-stellar mass space are presented for SFGs only. The three white lines in each panel represent the contours of log($M_*$/M$_\odot$) $= 9$, log($M_*$/M$_\odot$) $= 10$, and log($M_*$/M$_\odot$) $= 11$ at $z=3$. The evolution with redshift is derived from \citet{Weaver2023}. }
\label{fig:morp_z_m}
\end{figure*}

Understanding the relationship between galaxy mass and morphology is crucial for unveiling the underlying processes that drive the formation and evolution of galaxies. As displayed in Figure \ref{fig:morp_z_m}, we divide the star-forming galaxies (SFGs) into a $3 \times 3$ grid based on stellar mass and redshift. We then obtain the median values of each morphological parameter in the rest-frame V-band and J-band for each grid.

Our findings indicate that for both massive galaxies (log($M_*$/M$_\odot$)$ > $11) and less massive galaxies (log($M_*$/M$_\odot$)$ < $10), the changes in their morphological indicators ($C$, $A$, $G$, $M_{\rm 20}$) during stellar mass growth from high to low redshift are smaller than those for intermediate-mass galaxies, particularly in the rest-frame J-band. However, for the indicators $A_{\rm O}$ and $D_{\rm O}$, noticeable changes in morphology primarily occur in low-mass and intermediate-mass galaxies ($10 <$ log($M_*$/M$_\odot$) $<$ 11). This suggests that the outer asymmetric structure of galaxies remains relatively unchanged once the stellar mass reaches log($M_*$/M$_\odot$) $\sim 11$, and galaxies tend to become more rounded and regular. Additionally, indicators such as $C$, $Gini$, and $M_{\rm 20}$ show rapid growth within a narrow range of stellar masses up to log($M_*$/M$_\odot$) $\sim 10.5$. This is also evident in Figures \ref{fig:Rest_V_evo} and \ref{fig:Rest_J_evo}, indicating that the morphology of galaxies traced by these indicators is primarily influenced by stellar mass. On the other hand, the indicators $A$, $A_{\rm O}$, and $D_{\rm O}$ are correlated with both stellar mass and redshift.

\section{Discussions} \label{sec:Discussion}

\subsection{What are the Primary Factors Influencing the Morphological Indicators?}
\label{Primary Factors Influencing the Morphological Parameters}

\begin{figure}[!t]
\centering
\includegraphics[width=1\columnwidth] {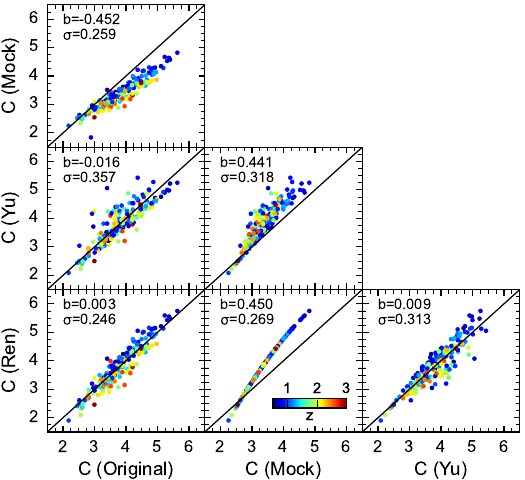} 
\caption{Comparison of the original concentration parameter ($C$) and the corrected concentration parameter ($C$) using our method and that of \citet{Yu2023} on TNG50 F200W images is presented.}
\label{fig:Compared_Yu}
\end{figure}

The impact of noise on different morphological indicators exhibits variability. Figure \ref{fig:Non-para_deviations} illustrates how noise within a specified range affects indicators such as $A$, $Gini$, and $A_{\rm O}$. However, since our \texttt{Mock} images only incorporate a single noise level, it is not possible to determine how the different morphological indicators change with increasing noise levels. While it would be valuable to understand the measurement biases of the morphological indicators under varying exposure depths and noise levels, such a discussion falls beyond the scope of the current study. It is important to note that although all our \texttt{Mock} images have the same noise level, galaxies with different surface brightness possess distinct signal-to-noise ratios ($S/N$). As \citet{Lotz2006} noted, morphological indicators measured in images with an $S/N$ lower than 2.5 should be cautiously treated. Therefore, our analysis of observed galaxy morphological indicators and their evolutionary trends is restricted to samples with $S/N > 2.5$.

The primary factor affecting the measurement of morphological indicators is the point spread function (PSF), and its influence primarily depends on the galaxy size. The impact of the PSF becomes more pronounced as its full width at half maximum (FWHM) approaches the size of the galaxy \citep{Wang2024, Yu2023}. This aspect was recognized during the PSF corrections performed by \citet{Yu2023}, where $r_{\rm p}$/FWHM was incorporated as a critical variable in the correction function. We selected samples from the TNG50 data with $\log(M_*/M_\odot) > 9.75$ and corrected the $C$ parameter for the F200W \texttt{Mock} images using the two methods. The results are shown in Figure \ref{fig:Compared_Yu}. Both correction methods are effective in mitigating the systematic bias affected by the PSF. However, our method exhibited a noticeably smaller scatter. This can be primarily attributed to the methodology proposed by \citet{Yu2023} involving two observational variables, observed $C$ and $r_{\rm p}$, resulting in a complex coupling of their dispersions and thus increasing the scatter following the correction. Studying the non-parametric morphological indicators of galaxies across a broad redshift range or in various filters using either uncorrected or PSF-matched morphological indicators can pose certain challenges, as demonstrated in Appendix B.

\subsection{Distributions of Different Components in Galaxies Indicated by Different Wavelengths}

Different components of galaxies contribute to the luminosity at different wavelengths. In the rest-frame UV band, the luminosity is primarily influenced by a combination of newly formed stars and dust obscuration. In contrast, the near-infrared (NIR) morphology is determined mainly by older stars, consistent with the distribution of stellar mass. Our findings reveal substantial variations in the different morphological indicators with wavelength in the optical band, consistent with previous studies \citep[e.g.][]{Baes2020, Nersesian2023, Yao2023}. \citet{Martorano2023} found, using a similar sample, that the radial profile distribution (S\'ersic index $n$) of galaxies is essentially invariant with wavelength. This suggests that the different components of galaxies roughly align along the radial profile in terms of overall distribution. However, in two-dimensional space, there is a significant difference in the distribution of the various components, which is reflected in the inhomogeneity of the distribution of newly formed massive stars. On the other hand, the distribution of low-mass stars tends to be more uniform and smooth.

\subsection{Morphological Evolution}

Although the current weak star formation rate (SFR) of quiescent galaxies (QGs) does not significantly impact their morphology, this study uncovers a strong correlation between red galaxy morphology and both stellar mass and redshift. This tendency aligns with the relaxation process experienced by galaxies, suggesting that the relaxation process significantly influences the morphological evolution of red galaxies. Notably, this process has been ongoing in red galaxies since $z=3$, but it becomes more prominent at $z<1.5$. It is worth emphasizing that a portion of low-redshift QGs has evolved from relatively higher-redshift star-forming galaxies (SFGs). These evolved QGs tend to exhibit higher stellar masses compared to QGs that have evolved from high-redshift sources, providing an additional crucial factor influencing the evolution of red galaxy morphology across different redshifts. As for SFGs, we propose that beyond a stellar mass threshold of $\log(M_*/M_\odot) \geq 10.5$, star formation activity begins to shut down, accompanied by the onset of the virialization process as galaxies transition into QGs. Some galaxies in this subset have already partially undergone this transformation, leading to a drastic metamorphosis in the morphology of SFGs above this mass threshold.

In conclusion, the interplay between redshift and stellar mass exerts a substantial influence on the morphology of galaxies, as these factors dictate the underlying physical processes occurring within galaxies.

\subsection{Merger Identification based on Non-parametric Morphological Indicators}

The identification of merging galaxy candidates is a widely used application of non-parametric morphological indicators. However, existing selection criteria, such as $C-A$ \citep{Conselice2003a}, $Gini-M_{\rm 20}$ \citep{Lotz2004, Lotz2008a}, and $A_{\rm O}-M_{\rm 20}$ \citep{Ren2023}, are primarily based on samples obtained at $z < 1$. As mentioned in this study, the point spread functions (PSFs) and detection depths of Hubble Space Telescope (HST) and James Webb Space Telescope (JWST) images differ, leading to systematic biases in measured morphological indicators. Additionally, non-parametric morphological indicators exhibit variations with redshift, rest-frame wavelength, and stellar mass. Therefore, conventional merger selection approaches based on non-parametric methods may introduce biases when applied to moderate to high redshift galaxies observed by {\it JWST}. While recent studies have examined the impact of different indicators on the selection of mergers using various indicators at different merger stages \citep{Ren2023, Rose2023, Wilkinson2024}, the current non-parametric based criterion for merger selection is no longer adequate for identifying high-redshift merging galaxies. Thus, there is a need to develop a revised criterion specifically tailored for the selection of high-redshift merging galaxies.

\begin{figure*}[!ht]
\centering
\includegraphics[width=1\textwidth] {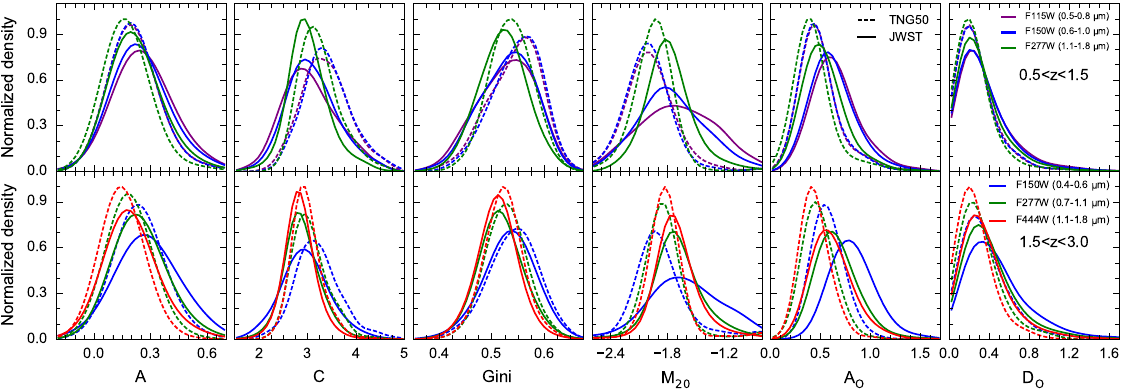} 
\caption{The distribution of non-parametric morphological indicators based on mock images of TNG50 galaxies and observed galaxies from {\it JWST}/NIRCam.}
\label{fig:TNG50_JWST}
\end{figure*}

\subsection{The Connection between Theories and Observations in Morphology}

We compare the morphological indicators distribution of TNG50 with observed galaxies from {\it JWST} in several filters in Figure \ref{fig:TNG50_JWST}. Our results demonstrate that at the longer wavelength filters, the morphology of TNG50 galaxies is generally consistent with that observed by {\it JWST}. However, there are significant differences at the shorter wavelength filters, especially in the $A_{\rm O}$ and $M_{\rm 20}$ indicators. The $A_{\rm O}$ and $M_{\rm 20}$ indicators for TNG50 galaxies are significantly smaller than those for observed galaxies at the short wavelength. This indicates a lack of some inhomogeneous substructures in the galaxies, determined by the methodology used by the Mock survey to generate the galaxy images, as shown in \citep{Snyder2023}. Some work uses the SKIRT code to consider radiative transfer to generate galaxies that are more consistent with observations \citep{Baes2024}, but these data are only available for massive galaxies in the local Universe.  If dust attenuation is considered in the galaxy images, then dust lanes or other substructures will appear in the shorter wavelength band, and these structures will make $A_{\rm O}$ and $M_{\rm 20}$ more significant, bringing the morphology closer to observations. On the other hand, if more star-forming clumps or fragmentation clumps appear in the distant TNG50 galaxies, then the morphology can be essentially consistent with observations at $z<3$.

The characterization and comparison of galaxy morphologies from observations and numerical simulations remain pivotal in understanding galaxy formation and evolution. The TNG50 simulation brings unprecedented resolution and physical complexity to the analysis of simulated galaxy morphologies \citep{Pillepich2019}. Comparisons between the morphologies of simulated galaxies in TNG50 and observed galaxies yield exciting insights. For instance, \citet{Rodriguez-Gomez2019} found that the morphological mix of galaxies in the TNG50 simulation successfully reproduces that of observed galaxies up to $z \sim 2$.

Several disparities also emerge between simulations and observations. At high redshifts ($z > 2$), TNG50 overproduces disc galaxies when compared with observations from the CANDELS fields \citep{Tacchella2019}. However, the latest observations from the {\it JWST} indicate that at these redshifts, there is indeed a higher proportion of disc galaxies in actual observations than those detected by the {\it HST} \citep{Ferreira2022, Ferreira2023, Kartaltepe2023, Kuhn2023}. This serves further to diminish the discrepancies between observational data and cosmological simulations.

\section{Conclusions}

Using a sample of more than 800 TNG50 galaxies with $\log(M_*/M_\odot) > 9$ at $0.5 < z < 3$, we investigated the differences in the morphological indicators of galaxies based on simulated images with Mock {\it JWST}/NIRCam observation conditions and those without observational effects. We then quantified the correlations and accordingly corrected for observational effects to the observed non-parametric indicators of over 4600 galaxies with $\log(M_*/M_\odot) > 9$ at $0.5 < z < 3$ in the {\it JWST} CEERS and JADES fields. Using the corrected morphological indicators, we studied the morphological evolution of SFGs and QGs. Our main results are as follows:

\begin{enumerate}

\item We uncovered that the noise level attributed to {\it JWST}/CEERS had negligible influence on the measurements of morphological indicators within the TNG50 sample. However, the impact of PSF played a significant role in accurately assessing morphological indicators, with a clear correlation between PSF effects and original morphological attributes. We employed second-order polynomials to effectively calibrate the morphological indicators derived from the TNG50 \texttt{Mock} images, which have similar stellar mass and redshift range as galaxies observed by {\it JWST}/NIRCam.

\item The variations of morphological indicators for star-forming and quiescent galaxies on rest-frame wavelengths have revealed intriguing findings. In the optical band, we observed a substantial evolution of morphological indicators with increasing wavelengths, with a pronounced trend at shorter wavelengths. This indicates that star-forming activity and extinction play crucial roles in shaping the galaxy morphology traced by non-parametric indicators. Conversely, in the near-infrared band, we observed minimal evolution of morphological indicators across the wavelength range. Furthermore, the indicators $A_{\rm O}$ and $M_{\rm 20}$ exhibit the most significant evolutionary trends, indicating their sensitivity to variations in star formation rate, extinction, and merger events.

\item We find that the evolution with redshift manifests more prominently in QGs than SFGs, with only indicators $A_{\rm O}$ and $D_{\rm O}$ exhibiting noticeable evolution. Based on this observation, we speculate that the increased occurrence of lopsided galaxies, attributed to galaxy mergers and disk instability, is largely influenced by the enhanced evolution witnessed in QGs at intermediate and high redshifts. For SFGs, the indicators $C$, $Gini$, and $M_{\rm 20}$ show a rapid evolution with stellar mass at $\log(M_*/M_\odot) \geq 10.5$, while $A_{\rm O}$, $D_{\rm O}$, and $A$ evolve with both redshift and stellar mass. This evolution can be ascribed to the formation and growth of the galactic bulges. Notably, the evolutionary trends in the optical band are more dominant than those in the near-infrared (NIR) band, but overall, the optical and NIR bands exhibit similar evolutionary patterns. The morphological differences of galaxies are smaller in the rest-frame J-band than in the rest-frame V-band. This suggests that the distribution of low-mass stars tends to be more uniform and smooth.

\item Through a comparison between the morphological indicators of the TNG50 \texttt{Mock} galaxies and those observed by {\it JWST}/NIRCam, we have identified a general agreement in terms of galaxy morphology. However, deviations are observed, particularly at the shorter wavelength for $A_{\rm O}$ and $M_{\rm 20}$. We attribute these discrepancies to the omission of dust in the TNG50 mock images, as well as the limited influence of extinction on the morphological structure at shorter wavelengths. To address this issue, the mock images should incorporate the effect of extinction to capture the true morphological characteristics accurately. Despite these limitations, the existing high-resolution hydrodynamic simulations demonstrate a significant level of consistency with observations regarding galaxy morphology.

\end{enumerate}

\section{acknowledgements}

We thank the anonymous referee for his or her expert and valuable comments, which have significantly improved this paper.  We also thank Zhigang An for his useful discussions. This project is supported by the National Natural Science Foundation of China (NSFC grants No. 12273052, 11733006) and the science research grants from the China Manned Space Project (No. CMS-CSST-2021-A04). NL acknowledges the support from the Ministry of Science and Technology of China (No. 2020SKA0110100), the science research grants from the China Manned Space Project (No. CMS-CSST-2021-A01), and the CAS Project for Young Scientists in Basic Research (No. YSBR-062). XZZ acknowledges the support from the National Science Foundation of China (Nos. 12233005 and 12073078) and the science research grants from the China Manned Space Project (Nos. CMS-CSST-2021-A02, CMS-CSST-2021-A04, and CMS-CSST-2021-A07). Some of the data presented in this paper were obtained from the Mikulski Archive for Space Telescopes (MAST) at the Space Telescope Science Institute. The JWST CEERS \citep{CEERS_data} and JADES \citep{JADES_data1, JADES_data2} data presented in this paper were obtained from the Mikulski Archive for Space Telescopes (MAST) at the Space Telescope Science Institute.

\bibliography{jwst_morphology}{}
\bibliographystyle{aasjournal}

\appendix

\section{The correction indices of non-parametric morphological indicators.}
\label{Correction index}
\begin{table*}[!ht]
\centering
\scriptsize
\caption{The correction parameters for non-parametric morphological indicators in the six {\it JWST}/NIRCam filters are as follows.}
\label{tab: pair samples}

\begin{tabular}{|c|c|c|c|c|c|c|c|c|c|c|c|c|c|c|c|c|c|c|c|c|}

\hline
\multicolumn{3}{|c}{\multirow{2}{*}{P}} &  \multicolumn{3}{|c}{F115W} & \multicolumn{3}{|c}{F150W} &  \multicolumn{3}{|c}{F200W} &  \multicolumn{3}{|c}{F277W} &  \multicolumn{3}{|c}{F356W} &  \multicolumn{3}{|c|}{F444W} \\ 

 \cline{4-21}

%\hline
\multicolumn{3}{|c|}{} &$\alpha$& $\beta$& $\gamma$ & $\alpha$& $\beta$& $\gamma$ & $\alpha$& $\beta$& $\gamma$ & $\alpha$& $\beta$& $\gamma$ & $\alpha$& $\beta$& $\gamma$ & $\alpha$& $\beta$& $\gamma$ \\
\hline
\multicolumn{3}{|c|}{$C$}& 0.04 & 1.05& -0.29 & -0.04& 1.54& -1.08 & -0.04& 1.58& 1.10 & 0.00& 1.37& -0.64 & -0.12& 2.26& -2.15 & -0.25 & 3.23& -3.79 \\
\hline
\multicolumn{3}{|c|}{$A$}& -1.18 & 2.05& 0.01 & -0.88& 1.81& 0.01 & -0.96&1.85& 0.03 & -0.87& 1.78& 0.06 & -1.02& 1.85& 0.05 & -1.22 & 1.97& 0.07 \\
\hline
\multicolumn{3}{|c|}{$Gini$}& 0.36 & 0.74& 0.06 & 0.01& 1.22& -0.10 & 0.34& 0.91& -0.03 & -1.55& 2.81& -0.49 & -0.34& 1.78& -0.27 & -0.90 & 2.64& -0.56 \\
\hline
\multicolumn{3}{|c|}{M$_{\rm 20}$}& -0.27& 0.36& -0.29 & -0.29& 0.31& -0.32 & -0.22& 0.60& -0.03 & -0.58& -0.42& -0.74 & -0.40& 0.28& -0.13 & -0.65 & -0.34& -0.45 \\
\hline
\multicolumn{3}{|c|}{A$_{\rm O}$}& -0.20 & 1.37& 0.04 & -0.11& 1.16& 0.08 & -0.20& 1.35& 0.03 & -0.13& 1.22& 0.08 & -0.19& 1.33& 0.02 & -0.42 & 1.68& -0.02 \\
\hline
\multicolumn{3}{|c|}{D$_{\rm O}$}& -0.27 & 1.48& -0.01& -0.17& 1.38& 0.01 & 0.18& 1.07& 0.03 & -0.08& 1.27& 0.03 & -0.28& 1.49& -0.01 & -0.35 & 1.71& -0.02 \\

\hline

\hline
 \end{tabular}
 \end{table*}

\newpage
\section{Comparison between the corrected and PSF-matched morphological indicators.}

\begin{figure*}[!h]
\centering
\includegraphics[width=1.0\textwidth] {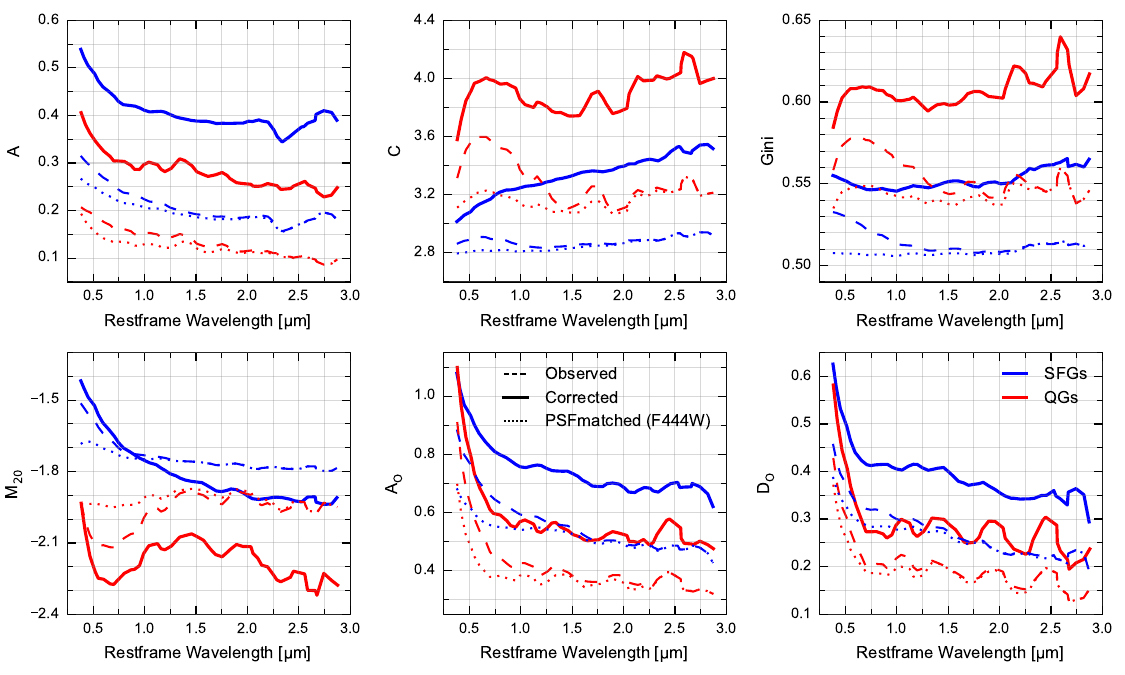} 
\caption{The wavelength dependence of non-parametric morphological indicators is shown. The solid, dashed, and dotted lines in each panel correspond to the corrected, observed, and PSF-matched to F444W indicators, respectively.}
\label{fig:Appendix_a}
\end{figure*}

A few studies have examined the evolution of non-parametric morphological indicators of galaxies by using the PSF matching technique to match all bands to the same PSF. Since the impact of the same PSF varies for different values of the same morphological parameter, we compare the evolution of the morphological parameter with wavelength for the morphological parameter corrected with our method, the morphological indicators obtained from direct measurements, and PSF matching to F444W. The results are presented in Figure \ref{fig:Appendix_a}. Our findings indicate that the directly measured and PSF-matched $A$, $A_{\rm O}$, and $D_{\rm O}$ exhibit a similar evolution trend with wavelength as the corrected indicators compared to the directly measured and PSF-matched $A$, $A_{\rm O}$, and $D_{\rm O}$. This suggests that the relationship between these three morphological indicators and wavelength is stronger than the effect of the PSF on them. However, the influence of the PSF on the indicators $C$, $Gini$, and $M_{\rm 20}$ is so significant in the optical bands that it even affects the trend of these indicators with wavelength. Thus, we emphasize that for PSF-sensitive indicators ($C$, $Gini$, and $M_{\rm 20}$), the impact of the PSF on the parameter values may outweigh the effect of wavelength.

\end{document}